\documentclass[article,lineno]{biometrika}
\usepackage{color,amsmath,amssymb,amsfonts,adjustbox}

\usepackage{times}
\usepackage{bm,verbatim}
\usepackage{natbib}
\usepackage[plain,noend]{algorithm2e}

\makeatletter
\renewcommand{\algocf@captiontext}[2]{\quad #1\algocf@typo. \AlCapFnt{}#2} 
\def\@algocf@capt@plain{top}
\renewcommand{\algocf@makecaption}[2]{%
  \addtolength{\hsize}{\algomargin}%
  \sbox\@tempboxa{\algocf@captiontext{#1}{#2}}%
  \ifdim\wd\@tempboxa >\hsize
    \hskip .5\algomargin%
    \parbox[t]{\hsize}{\algocf@captiontext{#1}{#2}}
  \else%
    \global\@minipagefalse%
    \hbox to\hsize{\box\@tempboxa}
  \fi%
  \addtolength{\hsize}{-\algomargin}%
}
\makeatother

\newcommand{\bX}{{X}}

\newcommand{\bz}{{ z}}

\newcommand{\mR}{\mathbb{R}}

\newcommand{\mS}{{\mathcal S}}

\newcommand{\var}{{\mbox{Var}}}

\newcommand{\bbeta}{{\beta}}
\newcommand{\btheta}{{ \theta}}

\newcommand{\bgamma}{{ \gamma}}

\begin{document}
\nolinenumbers
\jname{Biometrika}
\jyear{2019}
\jvol{}
\jnum{}



\markboth{Q. Song et~al.}{Biometrika style}

\title{Extended Stochastic Gradient MCMC for Large-Scale Bayesian Variable Selection}

\author{Qifan Song, Yan Sun, Mao Ye, Faming Liang}
\affil{Department of Statistics, Purdue University,\\ West lafayette, IN 47906, USA 
 \email{qfsong, sun748, ye207, fmliang@purdue.edu}}



\maketitle

\begin{abstract}
Stochastic gradient Markov chain Monte Carlo (MCMC) algorithms
have received much attention in Bayesian computing for big data problems, but they are only applicable to a small class of problems
for which the parameter space has a fixed dimension and the log-posterior density
is differentiable with respect to the parameters.
This paper proposes an extended stochastic gradient MCMC algorithm which, 
by introducing appropriate latent variables,
can be applied to more general large-scale Bayesian
computing problems, such as those involving dimension jumping and missing data.
Numerical studies show that the proposed algorithm is highly scalable and
much more efficient than traditional MCMC algorithms.
The proposed algorithms have much alleviated the pain of Bayesian methods
in big data computing.
\end{abstract}


\section{Introduction}

After six decades of continual development, MCMC has proven to be a powerful and
typically unique computational tool for analyzing data of complex structures.
However, for large datasets, 
its computational cost can be prohibitive as it requires all of 
the data to be processed at each iteration. 
To tackle this difficulty, a variety of scalable algorithms have been proposed in the recent literature. 
According to the strategies they employed, these algorithms can be grouped 
 into a few categories, including stochastic gradient MCMC algorithms 
\citep{Welling2011BayesianLV, Ding2014BayesianSU, Ahn2012BayesianPS, Chen2014StochasticGH, Betancourt2015,Ma2015ACR, NemethF2019}, 
split-and-merge algorithms  \citep{ConsensusMC2016, Srivastava2018, XueLiang2018},
mini-batch Metropolis-Hastings algorithms \citep{chen2016min,
korattikara2014austerity, bardenet2014towards, Maclaurin2014FireflyMC, Bardenet2017OnMC},
nonreversible Markov process-based algorithms \citep{BierkensFR2016,BouchardVD2016},
and some discrete sampling algorithms based on the multi-armed bandit \citep{ChenGhahramani2016}.   

Although scalable algorithms have been developed for both continuous and discrete sampling problems,
they are hard to be applied to dimension jumping problems. 
Dimension jumping is characterized by variable selection where 
the number of parameters changes from iteration to iteration in MCMC simulations. 
Under their current settings, the stochastic gradient MCMC and 
nonreversible Markov process-based algorithms are 
only applicable to problems for which the parameter space has a fixed dimension and
the log-posterior density is differentiable with respect to the parameters.
For the split-and-merge algorithms, it is unclear how to aggregate samples 
of different dimensions drawn from the posterior distributions 
based on different subset data. The multi-armed bandit algorithms are only applicable to 
problems with a small discrete domain and can be extremely inefficient for 
high-dimensional variable selection problems. 
The mini-batch Metropolis-Hastings algorithms are in principle applicable to 
dimension jumping problems. However, they are generally 
difficult to use. For example, the algorithms by \cite{chen2016min}, \cite{korattikara2014austerity}, 
and \cite{bardenet2014towards} perform approximate acceptance tests 
using subset data. The amount of data consumed for each test varies significantly 
from one iteration to another, which compromise their scalability.  
The algorithms by \cite{Maclaurin2014FireflyMC} and 
\cite{Bardenet2017OnMC} perform exact tests but require a lower bound on 
the parameter distribution across its domain. Unfortunately,
the lower bound is usually difficult to obtain.  

 This paper proposes an extended stochastic gradient 
 Langevin dynamics algorithm which, by introducing appropriate latent variables, 
 extends the stochastic gradient Langevin dynamics algorithm 
 to more general large-scale Bayesian computing problems 
 such as variable selection and missing data. The extended stochastic gradient 
 Langevin dynamics algorithm is highly scalable and much more efficient than traditional MCMC algorithms. 
 Compared to the mini-batch Metropolis-Hastings algorithms, 
 the proposed algorithm is much easier to use, which involves only a fixed amount of data 
 at each iteration and does not require any lower bound on the parameter distribution.
 

 \section{A Brief Review of Stochastic Gradient Langevin Dynamics} 

  Let $\bX_N=(X_1,X_2,\ldots$, $X_N)$ denote a set of $N$ independent and identically
  distributed samples drawn from the distribution $f(x|\btheta)$, where 
  $N$ is the sample size and $\btheta$ is the parameter. 
  Let $p(\bX_N|\btheta)=\prod_{i=1}^N f(X_i|\btheta)$ denote the likelihood function, 
  let $\pi(\btheta)$ denote the prior distribution of $\btheta$, and 
  let $\log \pi(\btheta|\bX_N)=\log p(\bX_N|\btheta) +\log \pi(\btheta)$ denote the
  log-posterior density function. If $\btheta$ has a fixed dimension and
  $\log\pi(\btheta|\bX_N)$ is differentiable with respect to $\btheta$, then the stochastic gradient
  Langevin dynamics algorithm \citep{Welling2011BayesianLV} can be applied to simulate from the posterior, which iterates by
  \begin{equation}\label{SGLDeq1} 
  \btheta_{t+1}=\btheta_t+\frac{\epsilon_{t+1}}{2} \widehat{\nabla}_{\btheta} \log\pi(\btheta_t|\bX_N) + 
            \surd(\epsilon_{t+1} \tau) \eta_{t+1},  \quad \eta_{t+1} \sim N(0,I_d), 
  \end{equation}
  where $d$ is the dimension of $\btheta$, $I_d$ is an $d\times d$-identity matrix,
  $\epsilon_{t+1}$ is the step size (also known as the
  learning rate), $\tau$ is the temperature,
  and $\widehat{\nabla}_{\btheta}\log \pi(\btheta_t|\bX_N)$ denotes an estimate of 
  $\nabla_{\btheta}\log \pi(\btheta_t| \bX_N)$ based on a mini-batch of samples. 
  The learning rate can be decreasing or kept as a constant. 
  For the former, the convergence of the algorithm was studied in 
   \cite{TehSGLD2016}.
  For the latter, the convergence of the algorithm was studied in \cite{Sato2014ApproximationAO} and \cite{DalalyanK2017}.  
 Refer to \cite{NemethF2019} for more discussions on the theory, implementation and variants of this algorithm.

 \section{An Extended Stochastic Gradient Langevin Dynamics Algorithm}
 
 To extend the applications of the stochastic gradient Langevin dynamics algorithm to varying-dimensional problems such as variable selection 
 and missing data, we first establish an identity for evaluating $\nabla_{\btheta} \log\pi(\btheta|\bX_N)$ in presence of latent variables. 
 As illustrated below, the latent variables can be the model indicator in the variable selection 
 problems or missing values in the missing data problems. 
 
 \begin{lemma} \label{lem0} For any latent variable $\vartheta$,    
 \begin{equation} \label{Fishereq1}
\nabla_{\btheta}\log\pi(\btheta \mid \bX_N)= 
 \int \nabla_{\btheta}\log\pi(\btheta \mid \vartheta,\bX_N) \pi(\vartheta \mid \btheta, \bX_N) d\vartheta,
\end{equation} 
where $\pi(\vartheta \mid \btheta, \bX_N)$ and $\pi(\btheta \mid \vartheta, \bX_N)$ denote the conditional 
distribution of $\vartheta$ and $\btheta$, respectively.
\end{lemma}

Lemma \ref{lem0} provides us a Monte Carlo estimator for $\nabla_{\btheta}\log\pi(\btheta \mid \bX_N)$ 
 by averaging over the samples drawn from the conditional distribution $\pi(\vartheta|\btheta,\bX_N)$. The identity (\ref{Fishereq1}) is similar to Fisher's identity. The latter has been used in evaluating the gradient of the log-likelihood function in presence of latent variables, see e.g. \cite{Cappe2005HMM}. 
 When $N$ is large, the computation can be accelerated by subsampling.
 Let $\bX_{n}$ denote a subsample, where $n$ denotes the subsample size. 
 Without loss of generality, we assume that $N$ is a multiple of $n$,
 i.e., $N/n$ is an integer. Let $\bX_{n,N}=\{X_n,\ldots, X_n\}$ denote a
 duplicated dataset with the subsample, whose total sample size is also $N$. 
 Following from (\ref{Fishereq1}), we have 
 \begin{equation} \label{dupeq}
 \nabla_{\btheta}\log\pi(\btheta \mid \bX_{n,N})
=\int \nabla_{\btheta}\log\pi(\btheta \mid \vartheta,\bX_{n,N}) \pi(\vartheta \mid \btheta, \bX_{n,N}) d\vartheta.
 \end{equation}
Since $\nabla_{\btheta}\log\pi(\btheta \mid \bX_{n,N})
=\nabla_{\btheta} \log p(\bX_{n,N}|\btheta)+\nabla_{\btheta} \log \pi(\btheta)$ is true  and $\log p(\bX_{n,N}|\btheta)$ is unbiased for $\log p(\bX_N|\btheta)$,  
$\nabla_{\btheta}\log\pi(\btheta \mid \bX_{n,N})$ forms 
an unbiased estimator of $\nabla_{\btheta}\log\pi(\btheta \mid \bX_{N})$.
 Sampling from $\pi(\bgamma_S|\btheta, \bX_{n,N})$ can be much faster than 
 sampling from
 $\pi(\bgamma_S|\btheta, \bX_{N})$ as for the former 
 the likelihood only needs to be 
 evaluated on a mini-batch of samples.

\subsection{Bayesian Variable Selection} 

 As an illustrative example, we consider the problem of variable selection for linear regression
 \begin{equation}\label{lm}
 Y = \bz^T \bbeta +\varepsilon,
\end{equation}
 where $\varepsilon$ is a zero-mean Gaussian random error with variance $\sigma^2$, 
 $\bbeta\in \mR^p$ is the vector of regression coefficients, 
 and $\bz=(\bz_1, \bz_2,\ldots, \bz_p)$ is the vector of explanatory variables. Let  
  $\bgamma_S=(\gamma_S^1,\ldots,\gamma_S^p)$ be a binary vector indicating the 
  variables included in model $S$, and let $\bbeta_S$ be the vector of 
  regression coefficients associated with the model $S$. From the perspective of Bayesian statistics, we are interested in estimating the posterior probability
  $\pi(\bgamma_S|\bX_N)$ for each model $S\in \mS$ 
  and the posterior mean $\pi(\rho)=\int \rho(\bbeta) \pi(\bbeta|\bX_N)$ 
   for some integrable function  $\rho(\cdot)$, where $\mathcal{S}$ comprises $2^p$ models. 
   Both quantities can be estimated using the reversible jump Metropolis-Hastings algorithm \citep{Green1995} by sampling from the posterior distribution $\pi(\bgamma_S,\bbeta_S|\bX_N)$. 
   However, when $N$ is large, the algorithm can be extremely slow due to 
   repeated scans of the full dataset in simulations. 
  
  As aforementioned, 
  the existing stochastic gradient MCMC algorithms cannot be  directly applied to simulate of   
  $\pi(\bgamma_S, \bbeta_S|\bX_N)$ due to the dimension jumping issue involved 
  in model transition. To address this issue, we introduce an auxiliary variable 
  $\btheta=(\theta^1,\theta^2,\ldots,\theta^p)$, 
  which links $\bgamma_S$ and $\bbeta_S$ through
 \begin{equation} \label{relationeq}
  \bbeta_{S}=\btheta \ast \bgamma_S=(\theta^1\gamma_S^1, \theta^2\gamma_S^2, \ldots, \theta^p\gamma_S^p),
  \end{equation}
 where $\ast$ denotes elementwise multiplication. Let 
 $\btheta_{[S]}=\{\theta^i: \gamma_S^i=1, i=1,2,\ldots,p\}$ 
  and $\btheta_{[-S]}=\{\theta^i: \gamma_S^i=0, i=1,2,\ldots,p\}$ 
 be subvectors of $\btheta$ corresponding to the nonzero and  zero elements of $\bgamma_S$, respectively. Note that $\bbeta_S$ is sparse with all elements in $\btheta_{[-S]}$ being zero, while $\btheta$ can be dense. Based on the relation (\ref{relationeq}), 
 we suggest to simulate from $\pi(\btheta|\bX_N)$ using the stochastic gradient Langevin 
 dynamic algorithm, for which the gradient $\nabla_{\btheta} \log\pi(\btheta|\bX_N)$ 
 can be evaluated using Lemma \ref{lem0}  by treating $\bgamma_S$ as the latent variable.
 Let $\pi(\btheta)$ denote the prior of $\btheta$. To simplify the 
  computation of $\nabla_{\btheta} \log\pi({\btheta}\mid \bgamma_S, \bX_N)$, we 
  further assume the {\it a priori} independence that
  $\pi(\btheta|\bgamma_S)=\pi(\btheta_{[S]}|\bgamma_S)\pi(\btheta_{[-S]}|\bgamma_S)$. 
  Then it is easy to derive 
 \[
 \nabla_{\btheta} \log\pi({\btheta}\mid \bgamma_S, \bX_N)= 
 \begin{cases}
 \nabla_{\btheta_{[S]}} \log p(\bX_N|\btheta_{[S]},\bgamma_S)+\nabla_{\btheta_{[S]}} \pi(\btheta_{[S]}|\bgamma_S),  
  & \mbox{for component $\btheta_{[S]}$,}\\   
 \nabla_{\btheta_{[-S]}} \log \pi(\btheta_{[-S]}|\bgamma_S), & \mbox{for component $\btheta_{[-S]}$,} \\ 
 \end{cases}
 \]
 which can be used in evaluating $\nabla \log\pi(\btheta|\bX_N)$ by Lemma \ref{lem0}. 
 If a mini-batch of data is used, the gradient can be evaluated based on (\ref{dupeq}). 
 This leads to an extended stochastic gradient langevin dynamics algorithm. 
  
 \begin{algo} \label{Alg2} [Extended Stochastic Gradient Langevin Dynamics for Bayesian Variable Selection]
  \begin{itemize}
   \item[(i)](Subsampling)  Draw a subsample of size $n$ (with or without replacement) 
      from the full dataset $\bX_N$ at random,
       and denote the subsample by $\bX_{n}^{(t)}$, where $t$ indexes the iteration.

   \item[(ii)] (Simulating models) Simulate models $\bgamma_{S_1,n}^{(t)},\ldots, \bgamma_{S_m,n}^{(t)}$ from 
     the conditional posterior $\pi(\bgamma_S |\btheta^{(t)},\bX_{n,N}^{(t)})$ by running a short Markov chain, where $\bX_{n,N}^{(t)}=\{X_n^{(t)},\ldots, X_n^{(t)}\}$ and $\btheta^{(t)}$ is the sample of $\btheta$ at iteration $t$.   

  \item[(iii)]  (Updating $\btheta$) Update $\btheta^{(t)}$ by setting
   $\btheta^{(t+1)} = \btheta^{(t)}+ (2m)^{-1}\epsilon_{t+1} 
     \sum_{k=1}^m \nabla_{\btheta}\log \pi(\btheta^{(t)}|\bgamma_{S_k,n}^{(t)}, \bX_{n,N}^{(t)})   
     + \surd(\epsilon_{t+1} \tau) \eta_{t+1}$,
   where $\epsilon_{t+1}$ is the learning rate, 
   $\eta_{t+1} \sim N(0,I_p)$, $\tau$ is the temperature, and $p$ is the dimension of $\btheta$. 
 \end{itemize}
 \end{algo} 
 
  Theorem \ref{thm:1} justifies the validity of this algorithm with the proof given in the Appendix.
\begin{theorem} \label{thm:1}
 Assume that the conditions (A.1)-(A.3) (given in Appendix) hold, $m$, $p$, $n$ are increasing with $N$ such that $N\geq n\succ p$, $m\succ {p}^{1/2}$, and a constant learning rate $\epsilon \prec{1}/{N}$ is used. Then, as $N\to\infty$,
 \begin{itemize}
     \item[(i)]  $W_2(\pi_t,\pi_*) \rightarrow 0$ as $t \to \infty$, where $\pi_t$ denotes the distribution of $\btheta^{(t)}$, $\pi_*=\pi(\btheta|\bX_N)$, and $W_2(\cdot,\cdot)$ denotes 
     the second order Wasserstein distance between two distributions. 
 \item[(ii)]  If $\rho(\btheta)$ is $\alpha$-Lipschitz for some constant $\alpha>0$, then $\sum_{t=1}^T \rho(\btheta^{(t)})/T \stackrel{p}{\to} \pi_*(\rho)$ as $T \to \infty$, where  $\stackrel{p}{\to}$ denotes convergence in probability and $\pi_*(\rho)=\int_{\Theta} \rho(\btheta) \pi(\btheta|\bX_N) d\btheta$.
 \item[(iii)] If (A.4) further holds, $\sum_{t=1}^T \sum_{i=1}^m I(\bgamma_{S_i,n}^{(t)}=\bgamma_S)/(mT)-\pi(\gamma_S|\bX_N) \stackrel{p}{\to} 0 $ 
 as $T\to\infty$.  
 \end{itemize}
\end{theorem}
 
 Part (i) establishes the weak convergence of $\btheta_t$; that is, if the total sample size $N$ and 
 the iteration number $t$ are sufficiently large, and the subsample size $n$ and the number of models $m$ simulated at each iteration are reasonably large, then $\pi(\btheta_t|\bX_N)$ will converge
 to the true posterior $\pi(\btheta|\bX_N)$ in 2-Wasserstein distance. Refer to \cite{GibbsSu2002} 
 for discussions on the relation between Wasserstein distance and other probability metrics. 
 Parts (ii) \& (iii) address our general interests on how to estimate the posterior mean and 
 posterior probability, respectively, based the samples simulated by Algorithm \ref{Alg2}. 
 For parts (i), (ii) and (iii), the explicit convergence rates are given in equations (\ref{W2eqcon}), (\ref{rate2}) and (\ref{rate3}), respectively.
 
For the choice of $m \succ p^{1/2}$, $p$ can be approximately treated as the maximum size of the models under consideration, which is of the same order as the true model. Therefore, $m$ can be pretty small under the model sparsity assumption.   
Theorem \ref{thm:1} is established with a constant learning rate. In practice, one may use a decaying learning rate, see e.g. \cite{TehSGLD2016},
 where it is suggested to set $\epsilon_t=O(1/t^{\kappa})$ for some $0<\kappa\leq 1$. 
For the decaying learning rate, \cite{TehSGLD2016} 
recommended some weighted averaging estimators for $\pi_*(\rho)$. Theorem \ref{thm:3} shows that the unweighted averaging estimators used above still work if the learning rate slowly decays at a rate of   
$\epsilon_t=O(1/t^{\kappa})$ for $0< \kappa<1$. However, if $\kappa=1$, the weighted averaging estimators are still needed. The proof of Theorem \ref{thm:3} is given in the supplementary material.  

\begin{theorem} \label{thm:3} Assume the conditions of Theorem \ref{thm:1} hold. If a decaying learning rate  $\epsilon_t=O(1/t^{\kappa})$ is used for some $0<\kappa<1$,  then parts (i), (ii) and (iii) 
 of Theorem \ref{thm:1} are still valid.
\end{theorem} 

\subsection{Missing Data}

 Missing data are ubiquitous over all fields from science to technology. However, 
 under the big data scenario, how to conduct Bayesian analysis in presence of missing 
 data is still unclear. The existing data-augmentation algorithm \citep{TannerWong1987} 
 is full data based and thus can be extremely slow. In this context,  
 we let $\bX_N$ denote the incomplete data and
 let $\btheta$ denote the model parameters.
 If we treat the missing values as latent 
 variables, then Lemma \ref{lem0} can be used for evaluating the gradient 
 $\nabla_{\btheta} \log \pi(\btheta|\bX_N)$.  
 However, Algorithm \ref{Alg2} cannot be directly applied to 
 missing data problems, since the imputation of  the missing data 
 might depend on the subsample only. 
 To address this issue, we propose Algorithm S1 (given in the Supplementary material), where 
 the missing values $\vartheta$ are imputed from $\pi(\vartheta|\btheta, \bX_{n})$ 
 at each iteration. 
 Theorem \ref{thm:1} and Theorem \ref{thm:3} are still applicable to this algorithm.

\section{An Illustrative Example}
This section illustrates the performance of Algorithm \ref{Alg2} using 
a simulated example. More numerical examples 
are presented in the supplementary material. 
 Ten synthetic datasets were generated from the model (\ref{lm}) with 
 $N=50,000$, $p=2001$, $\sigma^2=1$, $\beta_1=\cdots=\beta_5=1$,
$\beta_6=\beta_7=\beta_8=-1$, and $\beta_0=\beta_{9}=\cdots=\beta_p=0$, where $\sigma^2$ is assumed to be known, and the explanatory variables are normally distributed 
with a mutual correlation coefficient of 0.5.
A hierarchical prior was assumed for the model and parameters with the detail given in the supplementary material.
For each dataset, Algorithm \ref{Alg2} was run for 5000 iterations
with $n=200$, $m=10$, and 
the learning rate $\epsilon_t \equiv 10^{-6}$,  where the first 2000 iterations 
were discarded for the burn-in process and the samples generated from the 
remaining iterations were used for inference.  
 At each iteration, the reversible jump Metropolis-Hastings algorithm \citep{Green1995} was 
 used for simulating the models $\bgamma_{S_i,n}^{(t)}$, $i=1,2,\ldots,m$ with the detail given in
 the supplementary material.

Table \ref{boxtab2} summarizes the performance of the algorithm, 
 where the false selection rate (FSR), negative selection rate (NSR),  mean squared errors for false predictors ($\text{MSE}_0$) and mean squared errors for true predictors ($\text{MSE}_1$)
 are defined in the supplementary material.
The variables were selected according to the median posterior probability rule \citep{BarbieriBerger2004}, 
which selects only the variables with the marginal inclusion probability greater than 0.5. 
The Bayesian estimates of parameters were obtained by averaging over a 
set of thinned (by a factor of 10) posterior samples. 
 For comparison, some existing algorithms were applied to this example with 
 the results given in Table \ref{boxtab2} and the implementation details 
 given in the supplementary material. The comparison show that the proposed algorithm  
 has much alleviated the pain of Bayesian methods in 
 big data analysis. 
 
\begin{table}
\caption{Bayesian variable selection with the extended stochastic 
gradient Langevin dynamics (eSGLD), reversible jump Metropolis-Hastings (RJMH), 
split-and-merge (SaM) and Bayesian Lasso (B-Lasso) algorithms, 
where FSR, NSR, $\text{MSE}_1$ and $\text{MSE}_0$ are reported in averages 
 over 10 datasets with standard deviations given in the parentheses, 
 and the CPU time (in minutes) was 
 recorded for one dataset on a Linux machine with Intel\textsuperscript{\textregistered} Core\texttrademark i7-3770 CPU@3.40GHz.}
\label{boxtab2}
\begin{center}
\vspace{-0.2in}
\begin{adjustbox}{width=1\textwidth}
\begin{tabular}{cccccc} \hline
   Algorithm &  FSR &  NSR &  $\text{MSE}_1$ & $\text{MSE}_0$ & CPU(m) \\ \hline
   eSGLD     &  0(0)   & 0(0)    & $2.91\times 10^{-3}$($1.90\times 10^{-3}$) & $1.26\times 10^{-7}$($1.18\times 10^{-8}$) & 3.3 \\
       RJMH  &  0.50(0.10)  &  0.16(0.042)  & $1.60\times 10^{-1}$($3.89\times 10^{-2}$)   &  $2.64\times 10^{-5}$($8.75\times 10^{-6}$)   & 180.1 \\ 
       SaM &  0.05(0.05)  &  0.013(0.013)  & $1.29\times 10^{-2}$($1.27\times 10^{-2}$)   &  $1.01\times 10^{-6}$($1.00\times 10^{-6}$)   & 150.4 \\
  B-Lasso  & 0(0) &  0(0)  & $2.32\times 10^{-4}$($3.58\times 10^{-5}$)  &  $1.40\times 10^{-7}$($5.08\times 10^{-9}$)  & 32.8\\\hline 
\end{tabular}
\end{adjustbox}
\end{center}
\end{table}

{\centering \section{Discussion}}

 This paper has extended the stochastic gradient Langevin dynamics algorithm  
 to general large-scale Bayesian computing problems, 
 such as those involving dimension jumping and missing data. 
 To the best of our knowledge, this paper provides the first Bayesian method and theory for 
 high-dimensional discrete parameter estimation with mini-batch samples, while the existing methods
 work for continuous parameters or very low dimensional discrete problems only. 
 Other than generalized linear models, the proposed algorithm can
 have many applications in data science. For example, it can be used for 
 sparse deep learning   
 and accelerating computation 
 for statistical models/problems where 
 latent variables are involved, such as hidden Markov models, random coefficient models, and model-based clustering problems.


 Algorithm \ref{Alg2} can be further extended by updating $\btheta$ using 
 a variant of stochastic gradient Langevin dynamics, such as 
  stochastic gradient Hamiltonian Monte Carlo
 \citep{Chen2014StochasticGH}, 
 stochastic gradient thermostats \citep{Ding2014BayesianSU}, 
 stochastic gradient Fisher scoring \citep{Ahn2012BayesianPS},
  or preconditioned stochastic gradient Langevin dynamics \citep{Li2016PreconditionedSG}.  
We expect that the advantages of these variants  
(over stochastic gradient Langevin dynamics) 
can be carried over to the extension.

\vspace{2mm}

\section*{Acknowledgements}  
This work was partially supported by the grants DMS-1811812, DMS-1818674, and R01-GM126089. The authors thank the editor, associate editor and referees for their insightful comments/suggestions. 
\appendix
\section*{Appendix}

\subsection{Proof of Lemma \ref{lem0}}
\begin{proof}
 Let $\pi(\btheta)$ denote the prior density of $\btheta$, and let 
 $\pi(\vartheta)$ denote the density of $\vartheta$. Then 
 \[
 \begin{split}
&\nabla_{\btheta}\log\pi(\btheta | \bX_N)
=\nabla_{{\btheta}}\log p(\bX_N \mid {\btheta})+\nabla_{{\btheta}}\log\pi({\btheta}) 
=\frac{1}{p(\bX_N \mid{\btheta})}\nabla_{{\btheta}}\int  p(\bX_N,\vartheta 
  \mid\btheta) d\vartheta+\nabla_{{\btheta}}\log\pi({\btheta}) \\
&=\int \frac{p(\bX_N,\vartheta \mid{\btheta})}{p(\bX_N \mid{\btheta})}\nabla_{{\btheta}}\log p(\bX_N,\vartheta 
 \mid{\btheta})d\vartheta+\nabla_{{\btheta}}\log\pi({\btheta}) \\
&=\int \pi(\vartheta \mid{\btheta},\bX_N)\nabla_{{\btheta}}\left[\log 
 p(\bX_N \mid{\btheta},\vartheta)+\log\pi({\btheta}\mid\vartheta)
 +\log\pi(\vartheta)-\log\pi({\btheta})\right]d\vartheta +\nabla_{{\btheta}}\log\pi({\btheta})
\\
&=\int \nabla_{\btheta} \log\pi({\btheta}\mid\vartheta,\bX_N)  \pi(\vartheta \mid{\btheta},\bX_N) d\vartheta,  
 \end{split}
\]
where the second and third equalities follow from the relation $ \nabla_{\theta} \log(g(\theta))=\nabla_{\theta} g(\theta)/g(\theta)$ (for an appropriate function $g(\theta)$),
 and the fourth and fifth equalities are by direct calculations of the conditional 
 distributions.
\end{proof}

\subsection{Proof of Theorem \ref{thm:1}} 

Let $\pi_*=\pi(\btheta|\bX_N)$ denote the posterior density function of $\btheta$, and 
let $\pi_t = \pi(\btheta^{(t)}|\bX_N)$ denote the density 
of $\btheta^{(t)}$ generated by Algorithm \ref{Alg2} at iteration $t$.  
We are interested in studying the discrepancy between $\pi_*$ and $\pi_t$ in the 2nd order Wasserstein distance.
The following conditions are assumed. 
\begin{itemize}

\item[(A.1)] The posterior $\pi_*$ is strongly log-concave
and gradient-Lipschitz:
\begin{align} 
&f(\btheta)-f(\btheta^{\prime})-\nabla f(\btheta^{\prime})^T (\btheta-\btheta^{\prime}) \geq 
 \frac{q_N}{2} \|\btheta-\btheta^{\prime}\|_2^2,  \quad \forall \btheta,\btheta^{\prime} \in \Theta, \label{Aeq3}\\
 &\| \nabla f(\btheta)-\nabla f(\btheta^{\prime}) \|_2 \leq Q_N \|\btheta-\btheta^{\prime}\|_2, 
 \quad \forall \btheta, \btheta^{\prime} \in \Theta,   \label{A1eq2}
\end{align}
where $f(\btheta)= -\log \pi(\btheta|\bX_N)$, and $c_0'N\leq q_N\leq Q_N\leq c_0N$ for some positive constants $c_0$ and $c_0'$. 

\item[(A.2)] The posterior $\pi_*$ has bounded second moment: 
$\int_{\Theta}\btheta^T\btheta \pi_*(\btheta)d\btheta = O(p)$.

\end{itemize}

\begin{itemize}
\item[(A.3)]  
$ \max_{S\in \mS} E_{\bX_N}[\| \nabla_{\btheta}  \log \pi(\btheta|\bgamma_S,\bX_{N})\|^2|\btheta] =O(N^2(\|\theta\|^2+p))$, 
where $E_{\bX_N}$  denotes expectation with respect to the distribution of $\bX_N$, and $\mS$ denotes the set of all possible models.
 
\item[(A.4)] Let $L_N(\bgamma_S,\btheta)= \log p(X_N| \bgamma_S,\btheta)/N$ and let $\{L_N^{(i)}(\btheta): i=1,2,\ldots,|\mS|\}$ be the descending order statistics of $\{L_N(\bgamma_S,\btheta): S \in \mS\}$. Assume that there exists a constant $\delta>0$ such that $\inf_{\theta\in\Theta} (L_N^{(1)}(\theta)-L_{N}^{(2)}(\theta))\geq \delta$.
\end{itemize}

 \begin{proof} {\it Part (i).}
 In Algorithm \ref{Alg2}, the gradient $\nabla \log \pi(\btheta^{(t)}|\bX_{N})$ 
 is estimated by running a short Markov chain with a mini-batch of data.
 Since the initial distribution of the Markov chain might 
 not coincide with its equilibrium distribution, 
 the resulting gradient estimate can be biased. Let 
  $\zeta^{(t)}=\frac{1}{m} 
   \sum_{k=1}^m \nabla_{\btheta}\log \pi(\btheta^{(t)}|\bgamma_{S_k,n}^{(t)}, \bX_{n,N}^{(t)}) 
   -\nabla \log \pi(\btheta^{(t)}|\bX_{N})$.
 Following from (A.3), we have 
 \[
  \| E(\zeta^{(t)}| \btheta^{(t)})\|^2 =O\left[\frac{N^2(\|\btheta^{(t)}\|^2+p) }{m^2}\right], \,
  E\|\zeta^{(t)}- E(\zeta^{(t)}| \btheta^{(t)})\|^2  
   =O\left[\frac{N^2(\|\btheta^{(t)}\|^2+p)}{mn}\right].
 \]
  Following from Lemma S2 in the supplementary material, if $m \succ p^{1/2}$,  $\epsilon\prec {1}/{N} \prec ({mn})/({Np}),  $ and $V=O(p)$ holds, then 
  \begin{equation} \label{W2eqcon}
  W_2(\pi_t,\pi_*) =(1-\omega)^t W_2(\pi_0,\pi_*)+O\big(\frac{p^{1/2}}{m} \big) 
   +O((\epsilon p)^{1/2})+ O\big(\big(\frac{\epsilon Np}{mn}\big)^{1/2}\big)\to 0, \quad \mbox{as $t \to \infty$}, 
  \end{equation}
  for some $\omega>0$,
  since  $q_N\asymp N$ and $Q_N\asymp N$ hold by conditions (A.1) and (A.2). 
  
 {\it Part (ii).} Since $\rho(\btheta)$ is $\alpha$-Lipschitz, we have $|\rho(\theta)|\leq \alpha\|\theta\|+C'$ for some constant $C'$. Further, $\pi_*$ is strongly log-concave, so  $\pi_*(|\rho|) <\infty$, i.e., 
 $\rho$ is $\pi_*$-integrable.  On the other hand, 
\begin{equation}\label{W2}
\begin{split}
    &\|\int \rho(\btheta)d\pi_*(\btheta)-\int \rho(\tilde\theta)d \pi_t(\tilde\theta)\| = \|E \rho(\btheta)-E \rho(\tilde\theta)\| \leq E\|\rho(\btheta)-\rho(\tilde\theta)\|\\
    \leq&  \alpha E \|\btheta-\tilde\theta\|_2 \leq \alpha \{E\|\btheta-\tilde\theta\|_2^2\}^{1/2}= \alpha W_2(\pi_*,\pi_t)=o(1), \ \ 
     \mbox{ (due to eq. (\ref{W2eqcon})).}
\end{split}
\end{equation}
where $\btheta$ and $\tilde\theta$ are two random variables whose marginal distributions follow 
$\pi_*$ and $\pi_t$ respectively, $E(\cdot)$ denotes expectation with respect to the 
joint distribution of $\theta$ and $\tilde\theta$, and $(E\|\btheta-\theta_t\|_2^2)^{1/2}=W_2(\pi_*,\pi_t)$. 
This implies that $\rho$ is also $\pi_t$-integrable and 
$\int \rho(\tilde\theta)d \pi_t(\tilde\theta) \to \int \rho(\btheta)
d\pi_*(\btheta)$ as $t \to \infty$. 

Further, by the property of Markov chain, WLLN applies and thus 
$\sum_{t=1}^T \rho(\btheta^{(t)})/T 
-\sum_{t=1}^T \int \rho(\tilde\theta)d \pi_t(\tilde\theta)/T=O_p(T^{-1/2})$. Combining it with the above result leads to 
\begin{equation}\label{rate2}
\sum_{t=1}^T \rho(\btheta^{(t)})/T-\pi_*(\rho)=O_p(T^{-1/2})+\alpha \sum_{t=1}^TW_2(\pi_*,\pi_t)/T\rightarrow 0.
\end{equation} 
 
 {\it Part (iii).} To establish the convergence of $\hat{\pi}(\bgamma_S|\bX_N)$, we
  define $L_N(\bgamma_S,\btheta^{(t)})= \log p(\bX_N|\bgamma_S,\btheta^{(t)})/N$,
  $L_n(\bgamma_S,\btheta^{(t)})= \log p(\bX_n^{(t)}|\bgamma_S,\btheta^{(t)})/n$, and
  $\xi_{n,S}^{(t)}= L_n(\bgamma_S,\btheta^{(t)})- L_N(\bgamma_S,\btheta^{(t)})$ 
  for any $S \in \mS$.
  For each $S$, $\xi_{n,S}^{(t)}$ is approximately Gaussian  with $E(\xi_{n,S}^{(t)})=0$ and $\var(\xi_{n,S}^{(t)})=O(1/n)$.
  Therefore, for any positive $\nu$, with probability $1-|\mathcal S|^{-\nu}$, $\max_{S}|\xi_{n,S}|$ is bounded by $\delta_n:=\{(2\nu+2)\log|\mathcal S|/n\}^{1/2}=O[\{(\nu+1)p/n\}^{1/2}]$ according to the tail probability of the Gaussian. 
  It implies, with high probability, that  
if $S$ is the most likely model, i.e., $L_N^{(t)}(\gamma_S)=L_{N}^{(1)}(\theta^{(t)})$, then
\[
\begin{split}
&|\pi(\gamma_S|X_{n,N}^{(t)},\theta^{(t)})-\pi(\gamma_S|X_{N},\theta^{(t)})|\\
=& \big|\frac{1}{1+\sum_{S'\neq S}e^{N(L_N(\gamma_{S'},\theta^{(t)})-L_N(\gamma_S,\theta^{(t)})+\xi_{n,S'}-\xi_{n,S})}}
- \frac{1}{1+\sum_{S'\neq S}e^{N(L_N(X_N|\gamma_{S'},
\theta^{(t)})-L_N(X_N|\gamma_S,\theta^{(t)}))}}\big| \\
=& \frac{\sum_{S'\ne S} e^{N(L_N(X_N|\gamma_{S'},\theta^{(t)})-L_N(X_N|\gamma_S,\theta^{(t)}))+b_{S'})}}{[1+\sum_{S'\ne S} e^{N(L_N(X_N|\gamma_{S'},\theta^{(t)})-L_N(X_N|\gamma_S,\theta^{(t)})+b_{S'})}]^2}N |\xi_{n,S'}-\xi_{n,S}| \\ 
\leq &(2^p-1)e^{-N(\delta-2\delta_n)}N2\delta_n\leq e^{-N\delta/2}
\rightarrow 0,  
\end{split}
\]
if $\nu p\prec n$ (i.e., $\delta_n\prec \delta$) and $N\succ p$, 
where the second equality follows from the mean-value theorem by viewing $N(L_N(X_N|\gamma_{S'},\theta^{(t)})-L_N(X_N|\gamma_S,\theta^{(t)}))$'s as the arguments of  
$\pi(\bgamma_S|\bX_N,\btheta^{(t)})$, and $b_{S'}$ denotes a value between $0$ and $(\xi_{n,S'}-\xi_{n,S})$. 
Similarly, if $S$ is not the most likely model, then we denote $S^*$ as the most likely model and, by the mean-value theorem,  
\[
\begin{split}
&|\pi(\gamma_S|X_{n,N}^{(t)},\theta^{(t)})-\pi(\gamma_S|X_{N},\theta^{(t)})|\\
=& \bigg|\frac{e^{N(L_N(\gamma_S,\theta^{(t)})-L_N(\gamma_{S^*},\theta^{(t)})+\xi_{n,S}-\xi_{n,S^*})}}{1+\sum_{S'\neq S}e^{N(L_N(\gamma_{S'},\theta^{(t)})-L_N(\gamma_{S^*},\theta^{(t)})+\xi_{n,S'}-\xi_{n,{S^*}})}}
- \frac{e^{N(L_N(\gamma_S,\theta^{(t)})-L_N(\gamma_{S^*},\theta^{(t)}))}}{1+\sum_{S'\neq S}e^{N(L_N(\gamma_{S'},\theta^{(t)})-L_N(\gamma_{S^*},\theta^{(t)})}}\bigg|
\\
\leq &[1+(2^p-1)e^{-N(\delta-2\delta_n)}+e^{2N\delta_n}] e^{-N(\delta-2\delta_n)} N2\delta_n \leq e^{-N\delta/2}
\rightarrow 0.  
\end{split}
\]

In conclusion, with probability $1-1/|\mathcal S|^\nu$, 
$|\pi(\gamma_S|X_{n,N}^{(t)},\theta^{(t)})-\pi(\gamma_S|X_{N},\theta^{(t)})|<\exp(-N\delta/2)$ for all $S$, any iteration $t$ and any $\btheta^{(t)} \in \Theta$. Then, one could choose some $\nu = (n/p)^{1/2}\rightarrow\infty$, such that $\pi(\gamma_S|X_{n,N}^{(t)},\theta^{(t)})-\pi(\gamma_S|X_{N},\theta^{(t)})$  is bounded by  
\begin{equation} \label{biaseq}
\max_S E|\pi(\gamma_S|X_{n,N}^{(t)},\theta^{(t)})-\pi(\gamma_S|X_{N},\theta^{(t)})| \leq \exp(-N\delta/2)+1/|\mathcal S|^{\nu} \to 0,
\end{equation}
for any iteration $t$. 
Conditioned on $\{\btheta^{(t)}: t=1,2,\ldots\}$, $[\pi(\gamma_S|X_{n,N},\theta^{(t)})-\pi(\gamma_S|X_{N},\theta^{(t)})]$'s are independent and each 
 is bounded by 1, so WLLN applies.  
Therefore, for any $S\in \mS$, by WLLN, 
\begin{equation} \label{FMeq2}
 \frac{1}{T} \sum_{t=1}^T \pi(\bgamma_S |\bX_{n,N}^{(t)},\btheta^{(t)}) - 
 \frac{1}{T} \sum_{t=1}^T \pi(\bgamma_S |\bX_{N},\btheta^{(t)})
 =O_p(T^{-1/2})+\exp(-N\delta/2)+1/|\mathcal S|^{\nu}
 \rightarrow 0,
 \end{equation}
 provided  $p\prec n \leq N$. 
 Since $\{\btheta^{(t)}: t=1,2,\ldots\}$ forms a time-homogeneous Markov chain,
 whose convergence is measured by (\ref{W2eqcon}), and
 the function $\pi(\bgamma_S |\bX_{N},\btheta)$ is bounded and continuous in  $\btheta$,
 \begin{equation} \label{thetaeq}
 \frac{1}{T} \sum_{t=1}^T \pi(\bgamma_S|\bX_{N},\btheta^{(t)}) - 
 \pi(\gamma_S|\bX_N) =O_p(T^{-1/2}),
 \end{equation}
 holds for any $S \in \mS$. Combining (\ref{thetaeq}) with (\ref{FMeq2}) leads to
 \begin{equation} \label{FMeq3}
 \frac{1}{T} \sum_{t=1}^T \pi(\bgamma_S |\bX_{n,N}^{(t)},\btheta^{(t)})-\pi(\gamma_S|\bX_N) =O_p(T^{-1/2})+\exp(-N\delta/2)+1/|\mathcal S|^{\nu}
 \rightarrow 0.
 \end{equation}

  Conditioned on $\bX_{n,N}^{(t)}$ and $\btheta^{(t)}$, by the standard theory of MCMC, 
  $m^{-1} \sum_{i=1}^m I(\bgamma_S^{(t,i)}=\bgamma_S)$ forms a consistent estimator of 
  $\pi(\bgamma_S|\bX_{n,N}^{(t)},\btheta^{(t)})$ with an asymptotic bias of $O(1/m)$. Since $m$ is increasing with $p$ and $N$, the estimator is asymptotically unbiased. 
  Combining this result with (\ref{FMeq3}) leads to 
  \begin{equation}\label{rate3}
  \frac{1}{mT} \sum_{t=1}^T \sum_{i=1}^m I(\bgamma_{S_i,n}^{(t)}=\bgamma_S)-\pi(\gamma_S|\bX_N)=
   O_p(T^{-1/2})+\exp(-N\delta/2)+1/|\mathcal S|^{\nu}+O_p(m^{-1/2}),
   \end{equation}
   which converges to 0 as $T\to\infty$ and $N\to \infty$. 
\end{proof}

 
\bibliographystyle{biometrika}
\bibliography{ref}

\end{document}


\nolinenumbers

\setcounter{table}{0}
\renewcommand{\thetable}{S\arabic{table}}
\setcounter{figure}{0}
\renewcommand{\thefigure}{S\arabic{figure}}
\setcounter{equation}{0}
\renewcommand{\theequation}{S\arabic{equation}}
\setcounter{algo}{0}
\renewcommand{\thealgo}{S\arabic{algo}}
\setcounter{lemma}{0}
\renewcommand{\thelemma}{S\arabic{lemma}}

\jname{Biometrika}
\jyear{2019}
\jvol{}
\jnum{}



\markboth{Q. Song et~al.}{Biometrika style}

\title{Supplementary Material for ``Extended Stochastic Gradient MCMC for Large-Scale Bayesian Variable Selection''}

\author{Qifan Song, Yan Sun, Mao Ye, Faming Liang}
\affil{Department of Statistics, Purdue University,\\ West lafayette, IN 47906, USA 
 \email{qfsong, sun748, ye207, fmliang@purdue.edu}}



\maketitle

This material is organized as follows. Section \ref{sect1} presents an extended stochastic 
gradient Langevin dynamics algorithm for missing data problems. 
Section \ref{sect3} presents some lemmas and and the proof of Theorem 2. 
Section \ref{sect2} presents the reversible jump Metropolis-Hastings (RJMH) algorithm 
used in Algorithm 1 (of the main paper) for Bayesian variable selection.
Section \ref{sect4} presents more numerical examples.  

\section{An Extended Stochastic Gradient Langevin Dynamics Algorithm 
for Missing Data Problems} \label{sect1}
 
 As mentioned in the main paper, Algorithm 1 cannot be directly applied to  
 missing data problems, because the imputation of the missing data $\vartheta$ 
 might depend on the subsamples only. 
 To address this issue, we propose Algorithm \ref{Alg3}, where the missing values (denoted by 
 $\vartheta$) are imputed from $\pi(\vartheta|\btheta^{(t)}, \bX_{n})$ at each iteration $t$. 

  \begin{algo} \label{Alg3} [Extended Stochastic Gradient Langevin Dynamics for missing data problems]
  \begin{itemize}
   \item[(i)](Subsampling)  Draw a subsample of size $n$ from the full dataset $\bX_N$ at random,
              and denote the subsample by $\bX_{n}^{(t)}$, where $t$ indexes the iteration.

   \item[(ii)] (Multiple Imputation) For the subsamples, simulate the missing values $\vartheta_{1,n}^{(t)},\ldots, \vartheta_{m,n}^{(t)}$ from 
     the conditional posterior $\pi(\vartheta |\btheta^{(t)},\bX_n^{(t)})$,
     where $\btheta^{(t)}$ is the sample of $\btheta$ at iteration $t$.

  \item[(iii)] (Importance Resampling) Resample an imputed value
     from the set $\{\vartheta_{1,n}^{(t)},\ldots$, $\vartheta_{m,n}^{(t)}\}$ according to the
    importance weights $\{p(X_n|\btheta^{(t)}, \vartheta_{1,n}^{(t)})^{N/n-1}, \ldots, 
     p(X_n|\btheta^{(t)}, \vartheta_{m,n}^{(t)})^{N/n-1} \}$. Denote the resampled value by
     $\vartheta^{(t)}$.

  \item[(iv)] (Updating $\btheta$) Update $\btheta^{(t)}$ by setting
   \[
    \btheta^{(t+1)} = \btheta^{(t)}+ \frac{\epsilon_{t+1}}{2} \left\{\frac{N}{mn} 
     \sum_{k=1}^m \nabla_{\btheta}\log \pi(\btheta^{(t)}|\vartheta_{k,n}^{(t)}, \bX_{n}^{(t)})- \frac{N-n}{n}  
    \nabla_{\btheta} \log \pi(\btheta^{(t)})\right\} + \surd(\epsilon_{t+1} \tau) \eta_{t+1},
   \]
   where $\eta_{t+1} \sim N(0,I_d)$, and $\epsilon_{t+1}$ is the learning rate. 
  \end{itemize}
 \end{algo}
 
 In the case that $\vartheta$ or some components of $\vartheta$ depend on all data $\bX_N$, 
 an importance resampling step can be included 
 for simulating samples from $\pi(\vartheta|\btheta^{(t)}, \bX_{n,N})$ and then the corresponding 
 inference can be made. 
 For example, if Algorithm \ref{Alg3} is used for Bayesian 
 variable selection where the model is treated as latent variable,
 such a step is needed.  For the problems that $\vartheta$ depends on the 
  subsamples only, this step can be omitted. 
 The validity of Algorithm \ref{Alg3} follows directly from the fact that
  $N/n \nabla_{\btheta} \log \pi(\btheta|\bX_n)-(N-n)/n \nabla_{\btheta} 
 \log\pi(\btheta)$ is an unbiased estimator of $\nabla_{\btheta} \log \pi(\btheta|\bX_N)$.

 \section{Lemmas and Proof of Theorem 2} \label{sect3}
 
 \subsection{Some Lemmas}
 
\begin{lemma}\label{w20}
Let $X\sim\mu$ and $Y\sim\nu$, then \[E\|Y\|_2^2\leq E\|X\|_2^2+W_2^2(\mu,\nu)+2W_2(\mu,\nu)\{E\|X\|_2^2\}^{1/2},\]
 where $W_2(\cdot,\cdot)$ denotes the second order Wasserstein distance. 
\end{lemma}
\begin{proof}
By definition of the Wasserstein metric, without loss of generality, we assume that
$X$ and $Y$ satisfy $E\|X-Y\|_2^2=W_2^2(\mu,\nu)$. Then
\[
\begin{split}
    &[E\|Y\|_2^2-E\|X\|_2^2]-W_2^2(\mu,\nu)\\
    =&EY^TY-EX^TX-EX^TX-EY^TY+2EX^TY\\
    =&2EX^TY-2EX^TX = 2EX^T(Y-X)\leq 2E\|X\|_2\|Y-X\|_2\\
    \leq&2\{E\|X\|_2^2E\|Y-X\|_2^2\}^{1/2}=2W_2(\mu,\nu)\{E\|X\|_2^2\}^{1/2}.
\end{split}
\]

\end{proof}

Lemma \ref{lem2} is a generalization of Theorem 4 of \cite{DalalyanK2017}.
Compared to Theorem 4 of \cite{DalalyanK2017}, Lemma \ref{lem2} allows the noise of the stochastic gradients to scale with $\|\btheta^{(t-1)}\|^2$. A similar result can be found in \cite{bhatia2019bayesian}.

\begin{lemma} \label{lem2} 
 Let $\btheta^{(t-1)}$ and $\btheta^{(t)}$ be two random vectors in $\Theta$ satisfying
\[
 \btheta^{(t)} = \btheta^{(t-1)}-\epsilon [\nabla f(\btheta^{(t-1)})+\zeta^{(t-1)}]+\surd(2\epsilon)e^{(t)},
\]
where $\zeta^{(t-1)}$ denotes the independent random error of the gradient estimate which can depend on $\btheta^{(t-1)}$, and $e^{(t)} \sim N(0,I_p)$.   
Let $\pi_t$ be the distribution of $\btheta^{(t)}$, and let $\pi_*\propto \exp\{-f\}$ be the target distribution.
If $\zeta^{(t)}$ satisfies
 \[
 \| E(\zeta^{(t-1)}|\btheta^{(t-1)}) \|^2  \leq \eta^2 (\surd p+\|\btheta^{(t-1)}\|)^2, 
 \quad E[ \| \zeta^{(t-1)}- E(\zeta^{(t-1)}| \btheta^{(t-1)}) \|^2 ] \leq \sigma^2 (p+\|\btheta^{(t-1)}\|^2),
 \]
 for some constants $\eta$ and $\sigma$ which might depend on $m,n,p$ and $N$, and $\zeta^{(t-1)}$'s are independent of $e^{(t)}$'s, and if the conditions (A.1) and (A.2)  (given in the main paper) are satisfied,
 $q_N>\eta+\surd(2)\sigma$
 and  the learning rate $\epsilon \leq \min\{2/(q_N+Q_N), 1/q_N)\}$, then 
  \begin{equation} \label{coneq}
 \begin{split}
 W_2(\pi_t,\pi_*) &\leq [1-(q_N-\eta-\surd(2)\sigma)\epsilon]^t W_2(\pi_0,\pi_*)+ \frac{\eta (\surd{p}+\surd V)}{q_N+\eta+\surd(2)\sigma} +
  1.65 \frac{Q_N\surd(\epsilon p)}{q_N+\eta+\surd(2)\sigma} \\ &+\frac{\surd(\epsilon)\sigma^2(p+2V)}{1.65 Q_Np^{1/2}+ \{q_N+\eta+\surd(2)\sigma\}^{1/2}\{\sigma^2(p+2V)\}^{1/2}},
\end{split}
 \end{equation}
 where $V=\int \|\btheta\|^2 \pi_*(\btheta)d\btheta$. 
\end{lemma}
\begin{proof}
By the same arguments as used in the proof of Theorem 4 of \cite{DalalyanK2017}, 
we have 
\[
\begin{split}
   W_2^2(\pi_{k+1},\pi_*)\leq&\left[(1-q_N\epsilon)W_2(\pi_{k},\pi_*)+1.65Q_N(\epsilon^3p)^{1/2}+\epsilon\eta\sqrt p+\epsilon\eta \{E\|\btheta^{(k)}\|^2\}^{1/2}\right]^2\\
   &+\epsilon^2\sigma^2p+\epsilon^2\sigma^2E\|\btheta^{(k)}\|^2.
\end{split}
\]
By Lemma \ref{w20}$, E\|\btheta^{(k)}\|^2\leq (W_2(\pi_{k},\pi_*)+\surd V)^2$,
we can derive that
\[
\begin{split}
   W_2^2(\pi_{k+1},\pi_*)\leq&\left[(1-q_N\epsilon)W_2(\pi_{k},\pi_*)+1.65Q_N(\epsilon^3p)^{1/2}+\epsilon\eta\surd p+\epsilon\eta (W_2(\pi_{k},\pi_*)+\surd V)\right]^2\\
   &+\epsilon^2\sigma^2p+\epsilon^2\sigma^2(W_2(\pi_{k},\pi_*)+\surd V)^2\\
   \leq &\left[(1-q_N\epsilon+\eta\epsilon)W_2(\pi_{k},\pi_*)+1.65Q_N(\epsilon^3p)^{1/2}+\epsilon\eta(\surd p+\surd V)\right]^2\\
   &+\epsilon^2\sigma^2(p+2V)+2\epsilon^2\sigma^2W_2^2(\pi_{k},\pi_*)\\
   \leq &\left[(1-q_N\epsilon+\eta\epsilon+\surd(2) \sigma\epsilon)W_2(\pi_{k},\pi_*)+1.65Q_N(\epsilon^3p)^{1/2}+\epsilon\eta(\surd p+\surd V)\right]^2\\
   &+\epsilon^2\sigma^2(p+2V).
\end{split}
\]
The proof can then be concluded by Lemma 1 of \cite{DalalyanK2017}. 
\end{proof}

\begin{lemma}\label{wlln0}
Consider a stochastic gradient Langevin dynamics algorithm,
\[
\btheta^{(t)}=\btheta^{(t-1)}+\frac{\epsilon_t}{2}[F(\btheta^{(t-1)})+\zeta^{(t-1)}]+\surd(\epsilon_t)e^{(t)}, \quad e^{(t)}\sim \mbox{Normal}(0,1),
\]
for some smooth function $F$, 
where the independent random error term $\zeta^{(t-1)}$ can depend on $\btheta^{(t-1)}$ and $E\zeta^{(t-1)}=0$.
If the learning rate satisfies
$\lim_t\epsilon_t=0$, $\sum_{t=1}^\infty\epsilon_t=\infty$,
$\sum_{t=1}^{\infty}|1/\epsilon_{t+1}-1/\epsilon_t|/t\leq \infty$, and 
$\sum (\epsilon_tt^2)^{-1}\leq \infty$, and furthermore, there exists a Lyapunov function $V(\theta)$, which tends to infinity as $\|\btheta\|\rightarrow \infty$, is twice differentiable with
bounded second derivatives and satisfies the following conditions:
\begin{align}
& \langle F(\btheta),\btheta\rangle\leq -r\|\btheta\|\quad \mbox{ for all }\|\btheta\|>M\mbox{ and some $r,M>0$}, \label{3c}\\
       & E\|\zeta^{(t-1)}\|^{2k}\leq V^k(\btheta^{(t-1)})\quad \mbox{ for some }k\geq 2, \label{4c}\\
       & \|\nabla V(\btheta)\|^2+\|F(\btheta)\|^2\leq V(\btheta), \label{5c}\\
       & \langle\nabla V(\btheta), F(\btheta)\rangle\leq -\alpha V(\btheta) +\beta \quad\mbox{ for some }\alpha,\beta>0, \label{6c}
\end{align}
then the following results hold:  
\begin{enumerate}
    \item There exists a unique invariant distribution, $\pi$, for the diffusion process $d\theta_t=(1/2)F(\theta_t)dt+dW_t$.
    \item For any function $h$ such that $|h|/V^{k'}$ is bounded for some 
$k'\leq k/2$,
\[
\lim\frac{1}{T}\sum_{i=1}^T h(\btheta^{(i)})= E_\pi h(\btheta), \mbox{ a.s.}
\]
\end{enumerate}
\end{lemma}
\begin{proof}
Part 1: Condition (\ref{3c}) ensures existence of the invariant distribution by
proposition 4.1 of \cite{douc2009subgeometric}.

Part 2: The convergence of the sample path average is due to Theorem 7 of \cite{teh2016consistency}.
Note that Theorem 7 of \cite{teh2016consistency} shows  
the convergence of the weighted average $\sum_{i=1}^T w_i\btheta^{(t)}/\sum_{i=1}^T w_i$ with $w_i\rightarrow 0$ and $F=\nabla f$ for some $f$, but its proof is applicable 
to our case  (with a general drift $F$ and $w_i\equiv 1$ for all $i$) 
as well.
\end{proof}

\subsection{Proof of Theorem 2}

\begin{proof}
\begin{itemize}
\item[(i)] Without loss of generality, we let $\epsilon_t=1/t^{\kappa}$ where $\kappa\in(0,1)$. 
Define $\chi=\kappa/(1-\kappa)$, and let $T_0$ be the index such that $\max\{q_N, (Q_N+q_N)/2\}\epsilon_{T_0}<1$.
Let $T_0\leq T_1< T_2<\dots$ be a sequence of integers satisfying
$(s_0)^{-\chi} \leq \epsilon_{T_1}<\epsilon_{T_{2}-1}\leq (s_0+1)^{-\chi}
\leq \epsilon_{T_{2}}<\epsilon_{T_{3}-1}\leq (s_0+2)^{-\chi}\leq\dots
$, where $s_0$ is some integer. It is easy to see that 
$T_{i+1}-T_{i}\approx (\chi/\kappa)(s_0+i)^{\chi/\kappa-1}\asymp T_{i+1}^{\kappa}$.

Therefore, by the similar arguments as used in Lemma \ref{lem2} and Theorem 1, let $N$ be a fixed but sufficiently  large value,
\begin{equation}\label{ee1}
\begin{split}
 W_2(\pi_{T_{i+1}},\pi_*) & \leq (1-(q_N-\eta-\surd(2) \sigma)\epsilon_{T_{i+1}})^{T_{i+1}-T_i} W_2(\pi_{T_{i}},\pi_*)+C_1(\frac{\surd{p}}{m})\\ 
 & +C_2\left(\surd p+\left\{\frac{Np}{mn}\right\}^{1/2}\right)\surd{\epsilon_{T_i}}, \\
 \end{split}
\end{equation}
for some constant $C_1$ and $C_2$, where 
$q_N-\eta-\surd(2)\sigma>0$.
For a fixed value of $N$,  $(1-(q_N-\eta-\surd(2) \sigma)\epsilon_{T_{i+1}})^{T_{i+1}-T_i}\leq
(1-(q_N-\eta-\surd(2) \sigma)(T_{i+1})^{-\kappa})^{T_{i+1}-T_i}$ converges to some positive constant less than 1 as $i\to \infty$, since  $T_{i+1}-T_{i}\asymp T_{i+1}^{\kappa}$.
If we further assume that $(1-(q_N-\eta-\surd(2) \sigma)(T_{i+1})^{-\kappa})^{T_{i+1}-T_i}\leq \omega<1$ for all $i$, 
then (\ref{ee1}) is reduced to
\[
 W_2(\pi_{T_{i+1}},\pi_*)\leq \omega W_2(\pi_{T_{i}},\pi_*)+C_1(\frac{\surd{p}}{m})+C_2\left(\surd p+\left\{\frac{Np}{mn}\right\}^{1/2}\right)\surd{\epsilon_{T_i}}.
\]
Iterating the above inequality, we have
\[\begin{split}
 & W_2(\pi_{T_K},\pi_*)\leq \omega^K W_2(\pi_{T_0},\pi_*)+\sum_{i=1}^{K-1}\omega^i C_1\frac{\surd{p}}{m}+C_2\left(\surd p+\left\{\frac{Np}{mn}\right\}^{1/2}\right)\sum_{i=1}^{K}\omega^{K-i}\surd{\epsilon_{T_i}}\\
 &\leq \omega^K W_2(\pi_{T_0},\pi_*)+\omega/(1-\omega) C_1\frac{\surd p}{m}+C_2\left(\surd p+\left\{\frac{Np}{mn}\right\}^{1/2}\right)\sum_{i=1}^{K}\omega^{K-i}(s_{0}+i-1)^{-\chi/2}\\
 &\rightarrow \omega/(1-\omega) C_1\surd{p}/m \quad(\mbox{as }K\rightarrow\infty)\\
  &\rightarrow 0, \quad(\mbox{as } N\rightarrow\infty, \surd{p}/m\rightarrow 0),
 \end{split}
\]
which concludes part (i).

\item[(ii)] The extended stochastic gradient Langevin dynamics algorithm 
can be expressed as 
\[
\begin{split}
 \btheta^{(t+1)} &= \btheta^{(t)}+ \frac{\epsilon_{t+1}}{2m} 
     \sum_{k=1}^m \nabla_{\btheta}\log \pi(\btheta^{(t)}|\bgamma_{S_k}^{(t)}, \bX_{n,N}^{(t)})   
     + \surd(\epsilon_{t+1}) \eta_{t+1},\\
     &=\btheta^{(t)}+ \frac{\epsilon_{t+1}}{2} 
      \nabla_{\btheta}\log \pi(\btheta^{(t)}| \bX_{N}^{(t)})+ \frac{\epsilon_{t+1}}{2} \xi(\theta^{(t)})+ \frac{\epsilon_{t+1}}{2} \zeta^{(t)}   
     + \surd(\epsilon_{t+1}) \eta_{t+1},
\end{split}
\]
where 
$\xi(\theta^{(t)})=E[\sum_{k=1}^m \nabla_{\btheta}\log \pi(\btheta^{(t)}|\bgamma_{S_k}^{(t)}, \bX_{n,N}^{(t)})/m]-\nabla_{\btheta}\log \pi(\btheta^{(t)}| \bX_{N})$
and $\zeta^{(t)}=\sum_{k=1}^m \nabla_{\btheta}\log \pi(\btheta^{(t)}|\bgamma_{S_k}^{(t)}, \bX_{n,N}^{(t)})/m-E[\sum_{k=1}^m \nabla_{\btheta}\log \pi(\btheta^{(t)}|\bgamma_{S_k}^{(t)}, \bX_{n,N}^{(t)})/m]$
denote the bias term and noise term, respectively. 
It has been shown in the proof of Theorem 1 that
\[
  \|\xi(\btheta)\|^2 =o[\frac{N^2(\|\btheta\|^2+p) }{p}], \, \quad 
  E(\|\zeta^{(t)}\|^2|\btheta^{(t)})
   =O(\frac{N^2(\|\btheta^{(t)}\|^2+p)}{mn}).
 \]
With the same notation as in Lemma \ref{wlln0}, $F(\btheta)=\nabla\log\pi(\btheta|\bX_N)+\xi(\btheta)$. 
Because $-\log\pi(\btheta|\bX_N)$
is $Q_N$-gradient Lipschitz continuous and $q_N$-strongly convex with $q_N\asymp Q_N\asymp N$, we have
\[
\langle F(\theta),\theta\rangle\leq -q_N\|\theta\|^2+\|\theta\|\|\nabla_{\theta=0}\log\pi(\theta|X_N)\|+o(\frac{N}{\surd{p}}(\|\btheta\|^2+\|\btheta\|)),
\]
which implies condition (\ref{3c}).
It is also easy to verify conditions (\ref{4c})-(\ref{6c}) by choosing $k=1$ and $V(\theta)=C(\|\theta\|^2+1)$ for some sufficiently large constant $C$.
Also, the choice $\epsilon_t\propto t^{-\kappa}$ with $\kappa\in(0,1)$ satisfies the conditions of Lemma \ref{wlln0}, and any Lipschitz continuous $\rho$
satisfies that $|\rho|/V$ is bounded. 

Putting all these together it implies,
with given $\bX_N$, $N$, $p$ and $m$, that the
diffusion process $d\theta^{(t)}=(1/2)F(\theta^{(t)})dt+dW^{(t)}$ 
has a unique invariant distribution $\pi_N$, and 
\begin{equation} \label{thetaSupeq}
\lim\frac{1}{T}\sum_{t=1}^T \rho(\btheta^{(t)})= E_{\pi_N} \rho(\btheta), \mbox{ a.s.}
\end{equation}
As implied by Part (i) of Theorem 1, with proper growth rates of $m$ and $n$, $\pi_N$ converges to $\pi_*$ in $W_2$-distance. Hence, 
\begin{equation} \label{thetaSupeq2}
E_{\pi_N} \rho(\btheta)\rightarrow E_{\pi_*}\rho(\btheta).
\end{equation}
That is, the unweighted estimator given in part (ii) of Theorem 1 is still valid if the learning rate $\epsilon_t=O(1/t^{\kappa})$ for some $0<\kappa<1$.

\item[(iii)] For the proof of part (iii) of Theorem 1 in the main paper, equation (8)  still holds by (\ref{thetaSupeq}) and (\ref{thetaSupeq2}), and the other parts of the proof hold independent of the setting of the learning rate. Therefore, the unweighted estimator given in part (iii) of Theorem 1 is still valid if we set the learning rate $\epsilon_t=O(1/t^{\kappa})$ for some $0<\kappa<1$.
This concludes the proof of Theorem 2.  
\end{itemize}
\end{proof}


\section{The Reversible Jump Metropolis-Hastings Algorithm Used in Algorithm 1} \label{sect2}

 For Algorithm 1, we employ the reversible jump MH algorithm~\citep{Green1995} to 
 draw the models $\bgamma_{S_i,n}^{(t)}$, $i=1,2,\ldots,m$, at each iteration. 
 The reversible jump MH algorithm consists of three types of moves, namely, birth, death and exchange.
 Let $\{w_i\}_{i=0}^p$ denote the weights assigned to intercept term and variable $\bz_i$ for $i=1,2,\ldots,p$.
 For linear models, we assign $w_0=\exp(-1)$, $w_i=\exp(|\rho(\by,\bz_i)|-1)$ for $i=1,\dots,p$, where $\rho(\by,\bz_i)$ denotes the correlation coefficient between $\by$ and $\bz_i$; for generalized linear models, let $d_i$ denote the deviance of the model $S_i$, which consists of variable $\bz_i$ and the intercept term only, and we assign $w_0=\exp(-\epsilon)$, $w_i=\exp\{-(d_i-\min_j d_j)/(5 \sigma_d)-\epsilon\}$ for all $i=1,\dots,p$, where $\epsilon=0.1$ and $\sigma_d$ denotes the standard deviation of $d_1, d_2, \ldots, d_p$.
 Let $S^{(t)}$ denote the set of variables included in the current model, and
 let $S_c^{(t)}$ denote its complementary set.
 The birth move is to randomly select a variable from
 $S_c^{(t)}$ (with a probability proportional to its weight) and add it to $S^{(t)}$.
 The death move is to randomly select a variable
 from $S^{(t)}$ (with a probability proportional to one minus its weight $w_i$) and remove it.
 The exchange move is to randomly select a variable from
 $S^{(t)}$ (with a probability proportional to one minus its weight) and
 replace it by another variable randomly selected from $S_c^{(t)}$ (with a probability proportional to its weight).
 The exchange move keeps the size of $S^{(t)}$ unchanged.
 In addition, when a variable is selected to be added to or removed from  $S^{(t)}$, the sign of
 the corresponding component of $\btheta$ will be randomized, i.e., altered with
 probability 0.5. This accommodates with the prior setting of $\btheta_{[-S]}$ in equation (7) of the main paper. 
 In simulations, we set the proposal probability
 $P(\mbox{birth}|S^{(t)})= P(\mbox{death}|S^{(t)})=P(\mbox{exchange}|S^{(t)})=1/3$
 if $0<|S^{(t)}|<q$, $P(\mbox{birth}|S^{(t)})=1$ if $|S^{(t)}|=0$,
 and $P(\mbox{death}|S^{(t)})=1$ if $|S^{(t)}|=q$. Note that in the reversible jump moves,
 no regression coefficients were updated. This is different from
 traditional reversible jump MH algorithms, where the regression coefficients
 of $\btheta_{S^{(t)}}$ might need to be updated accordingly.

\section{Numerical Examples} \label{sect4}

\subsection{A Linear Regression Example}

This section provides more details for the illustrative example reported in the main paper. 
The true model follows
\begin{equation} \label{regmodel1}
 \by=\bz_{1}+\bz_{2}+\bz_{3}+\bz_{4}+\bz_{5}-\bz_6-\bz_7-\bz_8+0\cdot \bz_9+\cdots+0\cdot \bz_{p}+\varepsilon,
 \end{equation}
 where $\varepsilon\sim N(0,I_N)$, and $I_N$ denotes an $N\times N$ identity matrix. 
 The explanatory variables $\bz_{1}, \ldots, \bz_{p}$ were generated from a multivariate  Gaussian distribution with a mutual correlation coefficient of 0.5.
 We generated 10 independent datasets from the model (\ref{regmodel1}) with $N=50,000$ and $p=2000$. 
 A hierarchical prior was assumed for the model and parameters. 
 For each model $S$, we set
 \begin{equation} \label{prioreq1}
  \pi(\bgamma_S) \propto \lambda^{|S|}(1-\lambda)^{p+1-|S|} I(|S|\leq q),
 \end{equation}
 where $|S|$ denotes the number of explanatory variables included in the model,  $q$ is set to 50,
 and $\lambda$ is set to $1/(p+1)^{1.1}$ by including the intercept term. Here the factor $p+1$ means that the intercept term
 has been included in the model as a special explanatory variable.
 Conditioned on $\bgamma_S$, we set
 \begin{equation} \label{prioreq2}
\btheta_{[S]} \sim N(0,\sigma_1^2 I_{|S|}), \quad \btheta_{[-S]} \sim N(0,\sigma_0^2 I_{p+1-|S|}), 
\end{equation}
 where $\btheta_{[S]}$ and $\btheta_{[-S]}$ are the auxiliary 
 variables corresponding to the variables included in and excluded by the model $S$,
 respectively; $\sigma_1^2$ is set to a constant of 25; and  $\sigma_0^2$ is set to a constant of 0.025.

Algorithm 1 was run for 5000 iterations with the subsample size $n=200$,
the number of models $m=10$ drawn at each iteration, 
and the learning rate $\epsilon_t \equiv 10^{-6}$, where the first 2000 iterations 
were discarded for the burn-in process.
In each iteration, the subset models were drawn using the reversible jump Metropolis-Hastings 
algorithm described in Section \ref{sect2} of this material. 
Regarding the choice of the learning rate, we note that Theorem 1 of the main paper 
suggests that it should be set in the order $o(1/N)$. 
In all our examples, we  did not tune the learning rate much, 
just setting it around $1/N$ while avoiding possible 
blowups of the simulation.  Refer to \cite{NemethF2019} for   discussions 
on this issue for the general stochastic gradient Langevin dynamics algorithm.

The variables were selected according to the median posterior probability rule \citep{BarbieriBerger2004}, 
which selects only the variables with the marginal inclusion probability greater than 0.5. 
Table 1 of the main paper summarizes the performance of the algorithm, where the  
false selection rate (FSR), negative selection rate (NSR), 
$\text{MSE}_0$ and $\text{MSE}_1$ are reported in averages over 10 independent dataset.  
To be more precise, we define 
 \[
 \text{FSR}=\frac{1}{10}\sum_{i=1}^{10} |\hat{S}_i \setminus S_*|/|\hat{S}_i|, \quad  
 \text{NSR}=\frac{1}{10}\sum_{i=1}^{10} |S_*\setminus \hat{S}_i|/|S_*|,
 \]
 where $S_*$ denotes the set of true variables, $\hat{S}_i$ denotes the set of 
 selected variables for dataset $i$, and $|\cdot|$ denotes the cardinality of a set. 
 The accuracy of the parameter estimates is measured by  
 \[
 \text{MSE}_0=\frac{1}{10} \sum_{i=1}^{10} \text{MSE}(\hat{\bbeta}_i^{(0)}), \quad   
 \text{MSE}_1=\frac{1}{10} \sum_{i=1}^{10} \text{MSE}(\hat{\bbeta}_i^{(1)}),
\]
where $i$ indexes datasets, and $\hat{\bbeta}_i^{(0)}$ and $\hat{\bbeta}_i^{(1)}$ 
denote the estimates of the coefficients of the false and true variables, respectively. 
The $\hat{\bbeta}_i$ is obtained by averaging over a set of thinned (by a factor of 10) posterior samples produced by Algorithm 1. 
The extremely small values of $\text{MSE}_0$ and $\text{MSE}_1$ and the zero values 
 of FSR and NSR indicate that all the simulations have converged to the true model.

For comparison, some existing algorithms, including the reversible jump Metropolis-Hastings 
algorithm \citep{Green1995}, split-and-merge \citep{SongL2015b}, and Bayesian Lasso \citep{ParkC2008}, were applied to this example. 
For this example, the reversible jump Metropolis-Hastings algorithm
was also run for 5000 iterations. 
Similar to the implementation of extended stochastic gradient Langevin dynamic algorithm, 
the first 2000 iterations were discarded for the burn-in process, 
the variables were selected according to the median posterior probability rule \citep{BarbieriBerger2004}, and the regression coefficients were estimated based 
on the samples collected from the last 3000 iterations (thinned by a factor of 10). 


For the split-and-merge algorithm, we first split the data into 5 equal subsets along the dimension of predictors; that is, each subset consists of 400 predictors 
and 50,000 observations. 
For each subset, we imposed a hierarchical prior on the model and parameters, where 
each predictor has a prior probability of $(1/401)^{1.1}$ being selected, and if selected, its coefficient follows a Gaussian distribution $N(0,5^2)$. For each subset, the reversible 
jump Metropolis-Hastings algorithm was run for 2500 iterations, where the first 1000 iterations 
were discarded for the burn-in process, and the predictors with the marginal 
 posterior inclusion probability greater than 0.3 were selected. 
 The predictors selected from each subset were then merged into one dataset.  
 For the merged dataset, we imposed the same hierarchical prior on the model and parameters, 
 and then ran the reversible jump Metropolis-Hastings algorithm 
 for 2500 iterations with the first 1000 iterations discarded for the burn-in process. 
 The final model was selected according to the median posterior probability rule, and the 
 parameters were estimated based on the samples collected from the last 1500 iterations. 
  
For Bayesian Lasso, we adopt the Laplace prior $\pi(\bbeta)\propto \exp(-\lambda\sum|\beta_j|)$ with $\lambda=\{2N\log(p)\}^{1/2}$. The Gibbs sampler was used to 
 simulate posterior samples, as described in \cite{ParkC2008}, 
 for 200 iterations. The predictors with the absolute value of the 
 coefficient estimate greater than 0.1 were selected as the ``true'' predictors, where the
 coefficient was estimated by averaging its posterior sample over the iterations.  
  
Among the four algorithms, the extended stochastic gradient Langevin dynamics algorithm cost
the least CPU time and perfectly selected all true variables. 
The reversible jump Metropolis-Hastings algorithm took on average about 3 CPU hours to 
complete 5000 iterations. However, the resulting Markov chain was not yet mixed well 
and yielded high FSR, NSR and MSE$_1$ values.  
It is worth to note that the CPU time cost by the reversible-jump Metropolis-Hastings algorithm has a high variety 
among the ten runs, ranging from 42 to 305 minutes.
This is due to the local trapping issue suffered by the reversible-jump Metropolis-Hastings algorithm; the CPU time can dramatically increase if the Markov chain is frequently trapped at large models.
By working on lower dimensional subset data, the split-and-merge algorithm significantly 
improves the performance of the reversible jump Metropolis-Hastings algorithm, but is still 
inferior to the extended stochastic gradient Langevin dynamics algorithm in both variable 
selection accuracy and computational efficiency. We note that 
the split-and-merge algorithm can be further accelerated by parallel computing, 
although not done here.
The Bayesian Lasso was also capable to perfectly recover the true model. 
For this algorithm, the major CPU cost is for computing at each iteration  
the inverse of an $p\times p$-matrix, which, unfortunately, is not scalable with respect to $N$. 
Although the algorithm was only run for 200 iterations, it still cost more 
CPU time than the extended stochastic gradient Langevin dynamics algorithm. 
It is worth noting that other than linear regression, the Bayesian Lasso algorithm 
will not be so efficient as the conjugate Gibbs updates will not be available anymore.

 \subsection{A Logistic Regression Example} \label{logissec}

Other than the linear regression example reported in the main paper, we also tested Algorithm 1 on large-scale logistic regression. We simulated 10 large datasets from the model
\begin{equation} 
\mbox{logit} P(Y_i=1)=\sum_{i=1}^p \beta_j z_{ij}, \quad i=1,2,\ldots, N, 
\end{equation}
where $N=50,000$, $p=2000$, $\beta_1=\cdots=\beta_5=1$,
$\beta_6=\beta_7=\beta_8=-1$, and $\beta_{9}=\cdots=\beta_p=0$.
The explanatory variables $\bz_i$'s, where $\bz_i=(z_{i1}, \ldots, z_{ip})^T$,
were generated such that they have a mutual correlation coefficient of 0.5.
The response variables were generated such that half of them have a value of 1
and the other half have a value of 0.

To conduct Bayesian analysis for the datasets, we adopted the same prior setting as
 in \cite{LiangSY2013}. For each model $S$, we set
 \begin{equation} \label{GLMprioreq1}
  \pi(\bgamma_S) \propto \lambda^{|S|}(1-\lambda)^{p+1-|S|} I(|S|\leq q),
 \end{equation}
 where $|S|$ denotes the number of explanatory variables included in the model,  $q$ is set to 500,
 $\lambda$ is set to $1/\{1+(p+1)^{\zeta} \surd(2 \pi)\}$, and $\zeta=0.5$. Similar to the linear regression case, 
  the factor $p+1$ means that 
 the intercept term has been included in the model as a special explanatory variable.
 Conditioned on $\bgamma_S$, we set
 \begin{equation} \label{GLMprioreq2}
\btheta_{[S]} \sim N(0,\sigma_S^2 I_{|S|}), \quad \btheta_{[-S]} \sim N(0,\sigma_0^2 I_{p+1-|S|}), 
\end{equation}
 where $\btheta_{[S]}$ and $\btheta_{[-S]}$ denote the auxiliary variables corresponding to the variables included in and excluded by the model $S$,
 respectively;  $\sigma_0^2$ is set to a constant of 0.025; and $\sigma_S^2=\exp\{C_0/|S|\}/(2\pi)$.
 For this example, we set $C_0=10$.
 As shown in \cite{LiangSY2013}, such a prior setting ensures the posterior consistency and variable selection consistency for high-dimensional generalized linear models, 
 even when $p$ is much larger than $N$.

For each dataset, Algorithm 1 was run for 5000 iterations with the subsample size $n=300$,
the learning rate $\epsilon_t \equiv 10^{-5}$, and the number of models
$m=10$ simulated at each iteration, where the first 2000 iterations
were discarded for the burn-in process and the samples generated from the
remaining iterations were used for inference.
At each iteration, the models were simulated using the same reversible jump 
 Metropolis-Hastings algorithm
 as described in Section \ref{sect2} of this material.
 
 

The results were summarized in Table \ref{boxtab2}.
For each dataset of this example, the extended stochastic gradient Langevin dynamics 
algorithm took only 2.9 CPU minutes on a personal computer
of 3.6GHz! The numerical results indicate again that the extended 
stochastic gradient Langevin dynamic algorithm has much alleviated the pain of
Bayesian methods in big data analysis.

\begin{table}
\caption{ Bayesian variable selection for logistic regression 
 with the extended stochastic gradient Langevin Dynamics (eSGLD) algorithm,
where FSR, NSR, $\text{MSE}_1$ and $\text{MSE}_0$ are reported in averages 
 over 10 datasets with standard deviations given in the parentheses, 
 and the CPU time (in minutes) was 
 recorded for one dataset on a Linux machine with Intel(R) Core(TM) i7-7700 CPU@3.60GHz.}
\label{boxtab2}
\begin{center}
\vspace{-0.2in}
\begin{adjustbox}{width=1\textwidth}
\begin{tabular}{ccccccc} \hline\hline
 Model &  Algorithm &  FSR &  NSR &  ${MSE}_1$ & ${MSE}_0$ & CPU(minutes) \\ \hline
Logistic & eSGLD & 0(0) & 0(0) & $2.37\times 10^{-2}$($2.88\times 10^{-3}$)& $2.70\times 10^{-4}$($1.40\times 10^{-4}$)& 2.9
 \\ \hline\hline
\end{tabular}
\end{adjustbox}
\end{center}
\end{table}

\subsection{A Pascal Challenge Dataset} \label{Sect5.1}

We considered the dataset {\it epsilon} with logistic regression. The dataset
is available at  {\tt https://www.csie.ntu.edu.tw/$\sim$cjlin/libsvmtools/datasets},
which has been used for Pascal large scale
learning challenge in 2008. The raw dataset consists of 500,000 observations and 2000 features, which
was split into two parts: 400,000 observations for training and 100,000 observations for testing.
The training part is feature-wisely normalized to mean zero and variance one and then observation-wisely
scaled to unit length. Using the scaling factors of the training part,
the testing part is processed in a similar way.

For this example, we applied the same prior settings as used in Section \ref{logissec} 
 of this material except that
$C_0$ was set to 2.5.  A small value of $C_0$ enhances the selection of a larger set of variables,
while keeping the regression coefficients of the selected variables relatively small.
Our numerical experience shows that this often improves the prediction performance for real
data problems, as for which the true model might deviate from the logistic regression and thus
need to be approximated with a large number of variables (but the effect of each variable
is small).  With this prior setting,
Algorithm 1 was run for the dataset for 10 times independently.
Each run consisted of 1000 iterations, where the first 500 iterations were
discarded for the burn-in process, and the samples drawn from the remaining
iterations were used for inference. In simulations, we set the subsample size $n=2000$,
the learning rate $\epsilon_t \equiv 5\times 10^{-4}$, and the number of simulated models
$m=10$ per iteration.

The numerical results were summarized in Table \ref{epsilontab}, where the
variables were selected according to the median posterior probability rule \citep{BarbieriBerger2004}.
To measure the quality of the selected models, we evaluated the prediction
accuracy of the selected models on the test dataset. For comparison,
we applied the Lasso algorithm \citep{Tibshirani1996}, which was implemented using the package
{\it glmnet}, to this example. For Lasso, we reported only the results with one value of
 the regularization parameter $\lambda$
 at which it selected about the same size of models as eSGLD.
 A $t$-test for the eSGLD prediction error for the
 hypothesis $H_0: \nu=0.1644$ versus $H_1: \nu\neq 0.1644$ shows a $p$-value of
 $6.014\times 10^{-7}$, where $\nu$ denotes the prediction error. Therefore, the models selected by
 eSGLD have significantly lower predication error than that selected by Lasso.
 Later, we have tried other values of $C_0$ such as 1, 5 and 10.
 Under each of them, the models selected by eSGLD have
 significantly lower prediction error than the same size models selected by Lasso.

 For a thorough comparison, we have also applied the SCAD \citep{FanL2001} and MCP \citep{Zhang2010} algorithms
 to this example, which both were implemented in the R-Package {\it SIS} \citep{saldana2018sis}. Unfortunately,
 they did not produce any results after running 24 CPU hours. This comparison also shows
 that Lasso penalty is computationally more attractive than the SCAD and MCP penalties, although
 they share similar theoretical properties.
 
\begin{table}
\caption{Numerical results for the dataset {\it epsilon}, where the results of the 
 extended stochastic gradient Langevin dynamics (eSGLD) algorithm 
 were averaged over 10 independent runs with the standard deviation reported in the parentheses,
 and the CPU time was
 recorded for a run on a Linux machine with Intel(R) Core(TM) i7-7700 CPU@3.60GHz.}
\label{epsilontab}
\begin{center}
\begin{tabular}{ccccc} \hline\hline
 Algorithm & Setting   &  Model size  & Prediction Error($\nu$)   &  CPU (minutes) \\ \hline
  eSGLD    & $C_0=2.5$ &   62.7(2.6)  & 0.1422 ($1.80\times 10^{-3}$)  &  17.4 \\
  Lasso    & $\lambda=0.0239$  & 63   & 0.1644                         &  21.2 \\
  MCP/SCAD  & ---           & ---    &  ---                      &  $>1440$ \\ \hline\hline
\end{tabular}
\end{center}
\end{table}

 This example shows again that the extended stochastic gradient Langevin dynamics algorithm
 has much alleviated the pain of Bayesian computing for big data
 problems. For this example, it has even about the same CPU cost as the Lasso algorithm.

\subsection{MovieLens Data} \label{MovieSection}

Other than variable selection, we applied the extended stochastic gradient Langevin dynamics
algorithm to a
large-scale missing data problem, estimating the mixed effect model
for the MovieLens 10M Dataset downloaded at
{\tt https://grouplens.org/datasets/movielens/}.
The dataset contains $N=10,000,054$ ratings of 10,681 movies by 71,567 users. The
ratings vary from 0.5 to 5 in an increment of 0.5. The whole dataset contains
information of the ratings, the time of the ratings, and the genre of the
movies (with 19 possible categories).

This dataset has been modeled by mixed effect models, see e.g.
 \cite{van2000fitting} and \cite{srivastava2018asynchronous}.
Let $N$ be the total number of users, and let
 $s_{i}$ denote the number of ratings of user $i$.
 Following \cite{srivastava2018asynchronous}, we set
\[
{y}_{i}={W}_{i}{\beta}+{Z}_{i}{b}_{i}+{e}_{i}, 
 \quad {b}_{i}\sim N_{q}(0,{\Sigma}),\ {\Sigma}=\sigma^{2}{D},\ e_{i}\sim N_{s_{i}}(0,\sigma^{2}{I}_{s_{i}}),
\]
where ${y}_{i}\in \mR^{s_{i}}$, ${W}_{i}\in \mR^{s_{i}\times p}$, ${\beta}\in \mR^{p}$,
 ${Z}_{i}\in \mR^{s_{i}\times q}$, ${b}_i\in \mR^{q}$ for $i=1,2,\dots,N$. 
Let $\btheta=(\bbeta,D,\sigma^2)$ denote the parameter vector of the model, and  
let $r_{ij}$ be the rating of user $i$ for movie $j$. The response
is defined as ${y}_{i}=(r_{ij_{1}},r_{ij_{2}},\dots,r_{ij_{s_i}})^{T}$.
Here we assume that $({y}_{i},{W}_{i},{Z}_{i}),\ i=1,2,\dots,N$
are independent and identically distributed random samples with varying dimensions, 
where ${W}_{i}$ denotes the collection of
the intercept, genre predictor, popularity predictor and previous
predictor (defined below), and ${Z}_{i}$ is the same with ${W}_{i}$.

\begin{itemize}
\item { Genera predictor}, which is a categorical variable for four
categories, namely `Action', `Children', `Comedy', and `Drama',
grouped from 19 movie genres.
The category `Action' consists of the action, adventure, fantasy, horror, sci-fi, and thriller genres.
The category `Children' consists of the animation and children geres. The category `Comedy'
consists of only the comedy genre. The category `Drama' consists of the crime, documentary,
drama, film-noir, musical, mystery, romance, war, and western genres.
We use effect coding to represent each category, i.e., using
 $(1,0,0),(0,1,0),(0,0,1),(-1,-1,-1)$
to represent Children, Comedy, Drama and Action, respectively. The
genre predictor of movie $j$ is defined as the average of all categories
that  movie $j$ belongs to. For example, if movie $j$ belongs to Fantasy,
Thriller and Musical genres, then its genre predictor is
 $(1/3)\left((1,0,0)+(1,0,0)+(0,1,0)\right)=(2/3,1/3,0)$.

\item { Popularity predictor}, which, for the rating $r_{ij}$, is defined
 as $\text{logit}\{(l_j+0.5)/(L_j+1.0)\},$ where $L_j$ is the number
 of recent ratings of movie $j$, and $l_j$ is the number of recent ratings
 of movie $j$ with score greater than 3. Here ``recent'' means
 30 or fewer most recent ratings.

\item { Previous predictor}, which, for the rating $r_{ij}$, is defined
 as 1 if user $i$ rated previous movie with score greater than 3 and
 0 otherwise.
\end{itemize}

As in \cite{srivastava2018asynchronous}, we treat the random effect ${b}_{i}$ as missing data and set the priors
\[
{\beta}|\sigma^{2}\sim N_{p}(0,\ \sigma^{2}{I}_{p}),\ \sigma^{2}\sim\frac{1}{\chi_{2}^{2}},\ D\sim\text{inverse-Wishart}(p+2,\ {I}_{q}).
\]
In each iteration of Algorithm \ref{Alg3}, we first randomly select a mini-batch of users
with the subsample size $n=500$. For
the selected users, we sample the random effect ${b}_{i}$
based on $p({b}_{i}|{y}_{i},{\beta},\sigma^{2},{D})$
(see Equation 3.6 in \cite{van2000fitting} for its analytic form)
and then update $\theta=({\beta},{D}, \sigma^{2})$
according to the rule of Langevin Dynamics.
We use the linear regression result for ${Y}={X}{\beta}+{e}$
 to initialize $\bbeta$ and $\sigma^{2}$
and use a diagonal matrix with all diagonal elements equal to 0.2
to initialize $D$.
Algorithm \ref{Alg3} was run for 4000 iterations.
The learning rate was set to $0.002/N$ for this example. 

To compare with other competitive Bayesian sampling algorithm, we also perform the data augmentation Gibbs sampler, which iteratively samples $\theta$ and ${b}$ from the full data conditional distribution $\pi({b}|\theta, X_N)$ and $\pi(\theta|{b}, X_N)$, respectively where $\bX_N$ denotes the whole data. The algorithm was run for 500 iterations. 

The convergence performance of the extended stochastic gradient Langevin dynamics algorithm and Gibbs sampler algorithm
can be assessed by the kernel Stein discrepancy (KSD) \citep{gorham2015measuring,gorham2017measuring}. 
For a set of samples $\{\theta^{(t)}\}_{t=1}^T$ generated by a Markov chain, 
the kernel Stein discrepancy
between the empirical distribution of $\theta^{(t)}$, denoted by $\pi_t$, and the target posterior distribution $\pi_*=\pi(\btheta|\bX_N)$ is defined as 
\[
\text{KSD}(\pi_t,\pi_*)= \sum_{j=1}^p\left\{\sum_{t,t' = 1}^T \frac{k_j^0(\btheta^{(t)},\btheta^{(t')})}{T^2}\right\}^{1/2},
\]
where the Stein kernel for $j \in \{1,2,\ldots,p\}$ is given by 
$k_j^0(\btheta,\btheta') = \nabla_{\btheta_j}U(\btheta)\nabla_{\btheta'_{j}}U(\btheta')k(\btheta, \btheta') + \nabla_{\btheta_{j}}U(\btheta)\nabla_{\btheta'_{j}}k(\btheta, \btheta') + \nabla_{\btheta'_{j}}U(\btheta')\nabla_{\btheta_{j}}k(\btheta, \btheta') + \nabla_{\btheta_{j}}\nabla_{\btheta'_{j}}k(\btheta, \btheta')$, $p$ is the dimension of $\theta$, 
 $U(\btheta)=\log\pi(\btheta|\bX_N)$, 
 and $k(\btheta, \btheta') = (1 + ||\btheta - \btheta' ||_2^2)^{-0.5}$ is the inverse multi-quadratic kernel. 
 In practice, KSD can also be 
 calculated on some components of $\theta$ for simplicity. In this case, $p$ will be  smaller than the dimension of $\theta$. 
 
\begin{figure}
\begin{centering}
\includegraphics[scale=0.4]{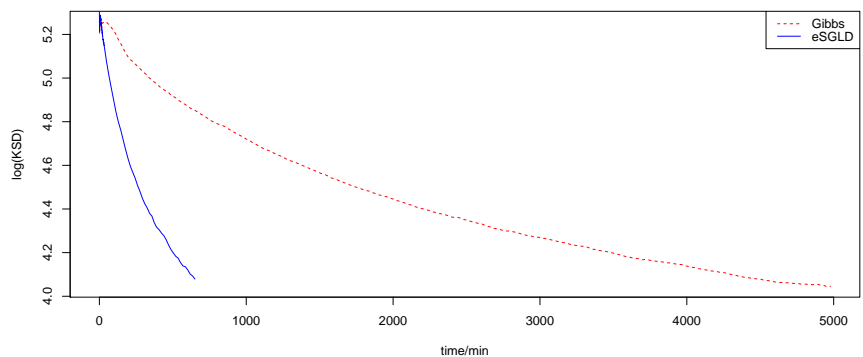}
\par\end{centering}
\caption{KSD ($\log_{10}$ scale) paths of $\beta$ against CPU time (minutes), where the solid line is for the  extended stochastic gradient Langevin dynamics (eSGLD) 
algorithm and the dotted line is for the Gibbs Sampler.}
\label{KSD_plot}
\end{figure}

\begin{figure}
\begin{centering}
\includegraphics[scale=0.45]{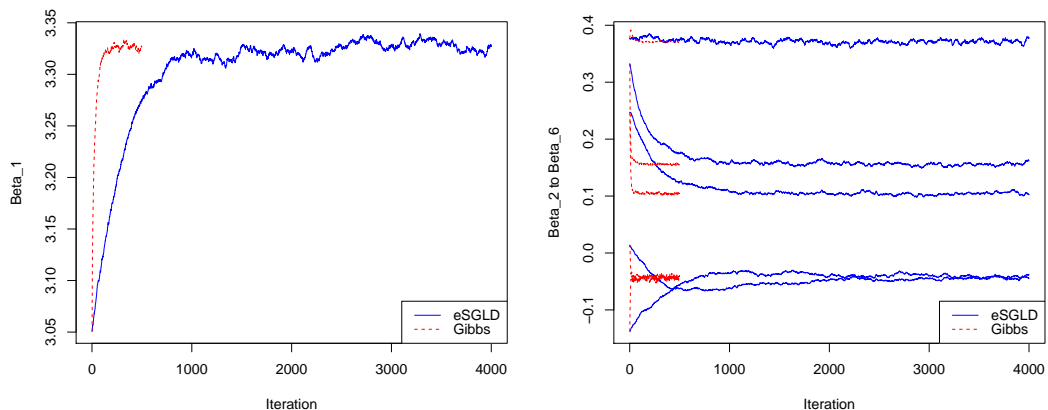}

\par\end{centering}
\caption{Sample paths of different components of $\bbeta=(\beta_1,\ldots,\beta_6)$, where
 the left panel is for $\beta_1$, and the right panel is for 
components $\beta_2$ to $\beta_6$.}
\label{Beta_path_plot}
\end{figure}

 Figure \ref{KSD_plot} compares the KSD paths of $\beta$  
 produced by the extended stochastic gradient Langevin dynamics algorithm and the Gibbs sampler, where the $x$-axis shows the elapsed CPU time. 
 Figure \ref{KSD_plot} indicates that to achieve the same level of KSD, the Gibbs sampler 
 costs much longer, approximately 7 times, CPU time than the 
 extended stochastic gradient Langevin dynamics algorithm. 
 Figure \ref{Beta_path_plot} plots the sample trajectories of $\beta_i$'s for both algorithms, with respect to the iteration number. It is easy to see that with same number of iterations, the extended stochastic gradient Langevin dynamics 
 algorithm does converge slower than the Gibbs sampler, however, such disadvantage 
 is compensated by its shorter CPU time cost per iteration.

\bibliographystyle{biometrika}
\bibliography{ref}